# Breaking Out of the Cell:
# On The Benefits of a New Spreadsheet User-Interaction Paradigm

Ziv Hellman
Inrise Financials Inc
ziv@inrisefinancials.com

**ABSTRACT**

*Contemporary spreadsheets are plagued by a profusion of errors, auditing difficulties, lack of uniform development methodologies, and barriers to easy comprehension of the underlying business models they represent. This paper presents a case that most of these difficulties stem from the fact that the standard spreadsheet user-interaction paradigm – the 'cell-matrix' approach – is appropriate for spreadsheet data presentation but has significant drawbacks with respect to spreadsheet creation, maintenance and comprehension when workbooks pass a minimal threshold of complexity. An alternative paradigm for the automated generation of spreadsheets directly from plain-language business model descriptions is presented along with its potential benefits. Sunsight Modeller$^{TM}$, a working software system implementing the suggested paradigm, is briefly described.*

## 1    INTRODUCTION

Even the most cursory perusal of journal literature on electronic spreadsheets in recent years will reveal that its dominant theme is consistently plaintive. The litany of dissatisfactions with the current state-of-affairs in the general use of spreadsheets includes the following:

1) **Profusion of spreadsheet errors**. There are a vast collection of studies on the prevalence of errors in spreadsheets [Panko, 1996]. Many studies [Chadwick, 2000; KPMG, 1997] consistently report that upwards of 90% of spreadsheets that pass a certain threshold of complexity may contain major errors (defined as errors that could affect decisions based on the results of the model), most of them due to individually seemingly trivial actions such as erroneous cut-and-paste, incorrect cell-references, obsolete absolute cell-references and structurally disruptive column-and-row alterations. EuSpRIG maintains an entire Web-page devoted to tracking press reports of spreadsheet errors. Given that the preponderant use of spreadsheets in professional activity is related to business and financial reporting and decision-making, the total cost of spreadsheet errors is enormous.

2) **Auditing difficulties**. A study by three PricewaterhouseCoopers researchers [Ettema et al, 2001] reports on the cost of auditing a typical spreadsheet in the following manner: The number of distinct formulae depends on the nature of the spreadsheet, but for moderately sized spreadsheets this number typically lies in the range of 500 to 1,500. The inspection of a distinct formula by an experienced auditor takes on average around 3 minutes. The total effort of a traditional spreadsheet audit generally takes from 25 up to 75 hours. If the number of distinct formulae is much larger 1,500, or if the spreadsheet is complicated and badly structured, it can be difficult to perform an audit by more than one auditor in parallel. The throughput time for auditing such spreadsheets will be, based on these assumptions, more than a fortnight. Given these numbers and typical corporate time pressures, spreadsheet audits are too often performed in a cursory manner, if at all. The attendant risks, especially in light of the requirements of the Sarbanes-Oxley act, are alarming

3) **Absence of Documentation**. The absence of documentation has been a factor in a number of well-documented spreadsheet errors [Butler, 2000]. Failure to document can lead to serious errors and maintenance nightmares, especially in environments where a model is passed from one user to another.

4) **Lack of Uniform Development Methodologies**. It appears that most organisations do not have even the most rudimentary internal modelling standards [Chadwick, 2000]. The lack of uniform corporate-wide standards, methodologies and presentation styles hinders quality control and multiplies confusion.

5) **Re-use is Nearly Non-existent.** [Ettema et al, 2001] report that PWC has collected a wide collection of formulae that appear over and over again in many spreadsheets. Efficiency interests would indicate that these formulae ought to be re-used from already existing spreadsheets by spreadsheet creators, rather then re-composed from scratch and re-audited in each new spreadsheet. This is not done because re-use typically involves cut-and-paste which is tedious and error-prone.

This paper presents the thesis that all of the above problems are related to aspects of the main spreadsheet interface paradigm – which may be termed the 'cell-matrix' paradigm – invented in the 1970's and propagated nearly unchanged to this day. The cell-matrix paradigm is excellent for the presentation of data in spreadsheets, but unwieldy and error-prone with respect to other actions such as spreadsheet creation, maintenance, model comprehension, and auditing. It follows that the corrective should involve the introduction of different interface paradigms. The paper ends with a brief presentation of Sunsight Modeller$^{TM}$, a commercially-oriented software package that implements in practice the ideas brought forth here.

## 2 AFFORDANCE AND PARADIGMS

### 2.1 Affordance

*Affordance* is a word originally coined in the psychological research of James Gibson, but its most widespread introduction came from books on the design of objects by Donald Norman [Norman, 1988]. The meaning of the word is perhaps best understood by example. A person approaches a door designed to appear as a monolithic slab. In which direction does it open? Should one pull or push, on the left or on the right? Perhaps it slides open? If, on the other hand, the door has on it a prominent flat handle with an illustration of a human palm, one would have no hesitation walking up to the handle, placing a hand against it, and pushing the door open. The handle 'affords' being pushed.

Based on such examples, a literature on affordance and its application to the design of everything from staplers to pavement paths has emerged. The basic principles of affordance, as stated in the ample literature [Gibson, 1977; Gibson 1979; Norman, 1988; Norman, 1990] can be summarised in a handful of sentences: Take advantage of analogies and cultural standards. Place a stress on human intuition, natural thought-processes, retention, prior experience, natural language, and consistency. Strive to make the most important features salient and obvious.

In the computing context, the idea of affordance has been most widely adopted by professionals working on graphic user interface design, and is often applied with respect to questions such as how a clickable button should appear and where, or in what context to use radio-buttons as opposed to tabs. It can, however, be considered in a more systemic context, as applying to an entire paradigm of interface. The most prominent example of a shift to what may be termed an *affordance paradigm* is the adoption of the desktop metaphor supplanting the older command-line paradigm. The history of the evolution of computer languages, from machine code to assembler to functional programming to object-oriented languages, may similarly be viewed as shifts of interface paradigm intended to boost affordance by ever more closely mimicking human intuitions and languages, and making central features salient.

**2.2　The Cell-Matrix Paradigm**

The commercial electronic spreadsheet was invented in the late 1970's by Daniel Bricklin and Robert Frankston. It represented a seminal event in the adoption of personal computers in businesses and homes, and its impact on modern business methods cannot be overstated. As the name suggests, the electronic spreadsheet took its metaphor from the realm of accounting, where a "spread sheet" meant a large sheet of paper with columns and rows that organised data about transactions for a business person to examine. Bricklin has also been quoted as saying he was influenced by blackboard presentations of data in business school lectures (cf. www.bricklin.com). These conceptual roots form the bedrock of the main spreadsheet interface paradigm, which may be termed the 'cell-matrix paradigm'. In its very familiar essence, particular cells within a matrix are designated to present data. Data in one cell is inter-related to data in another cell, thus ensuring that the entire matrix presents a conceptual whole, by attaching a formula to a cell, with that formula referring to other cells by code, typically by a combination of row and column designators such as C24 or H86, for example.

From the perspective solely of the dimension of data presentation, the cell-matrix paradigm is an excellent example of applying the principles of affordance in the field of computer automation, which is doubtless a major factor in its phenomenal success. By borrowing its presentation style from the spread-sheets used for centuries in businesses, it took full advantage of analogies, prior experience and cultural standards. Reading a spreadsheet for the purpose of seeing the numbers and taking in 'the bottom line' is intuitive for humans, and, albeit with some effort, spreadsheets can be styled to make their most important data appear to be the most salient features. The automation of calculation borne by the cell formulae frees humans from performing tedious and repetitious tasks.

The problems begin because the very same cell-matrix paradigm is also used for the purposes of spreadsheet creation, comprehension and auditing. It utilises the 'What You See is What You Get' interface, conflating structure, presentation and content. This type of interface is quite agreeable in many other situations but fails in the context of spreadsheets. It is a tolerable and perhaps even an inviting metaphor when one is creating a small spreadsheet, in a 'tinkering' manner. But when spreadsheets grow beyond a certain threshold of complexity, the strain of retaining in the mind scattered cryptic cell-references in hundreds or thousands of formulae becomes too much to bear. One may reference in this regard studies on the importance of distinctions between presentation, data and the underlying logic in the authoring of computer-related presentations [Isakowitz et al, 1995].

Human beings do not naturally think of business models in terms of cells in a matrix. 'Net Income' bears far more meaning than cell-reference C38. When schools of business administration and accountancy present business models, they do so using conceptually related collections of formulae in verbal human language. The mental effort required for translating between this model and the cell-matrix metaphor may be regarded as the root of many difficulties. Cell-references that are meaningless in and of themselves are an open invitation to errors. Filling in every cell in an immense matrix is a tedious, repetitive and time-consuming task, with copy-and-paste actions serving as a further pitfall for errors. Reconfiguring a matrix structure to fit an evolving mental model can similarly be tedious and error-prone. The complexity of large and multiple matrix workbooks coupled with the wearisome cell-chasing required for elucidating the meaning of formulae is an impediment to comprehension and drives up the costs of auditing. The traditional spreadsheet does not afford ample documentation. Reusability of formulae from one spreadsheet to another is rendered nearly impossible by the strictures of the cell-reference method. The design of the interface, which involves manual specification of structure, presentation and content, works against uniformity of presentation and hinders the adoption of corporate-wide methodologies.

Researchers studying the most effective methods for increasing spreadsheet accuracy [Kruck, Sheetz, 2001] generally recommend careful planning and design of spreadsheets, simplification of formula complexity and testing of spreadsheets. There is also ample empirical evidence that many spreadsheets are created in a collaborative manner rather than isolated individuals [Nardi, 1993]. All of these arguably are hindered by the above-listed drawbacks of the cell-matrix interface.

## 2.3    Out of the Shadows

The most obvious partial remedy to the above-presented situation is to use named references for cells. [Napier et al, 1989] compared the performance of novices using Lotus HAL with Lotus 1-2-3, and concluded that HAL users consistently solved more problems because the language more readily allowed reference to spread-sheet cells by names. Named references for cells have long been available in most commercial spreadsheet software products, and the intuition that consistent use of named references contributes significantly to reducing errors is also supported by the study [Janvarin and Whittle, 2003]. It appears, however, that most users of spreadsheets do not make use of this feature.

In the paper *Spreadsheet assurance by "control around" is a viable alternative to the traditional approach* [Ettema et al, 2001], a case is presented for utilising what is termed a 'shadow model' as an aid for auditing. The shadow model consists of formulae written in plain English which model the spreadsheet that is to be audited. The audit methodology then consists of importing scenarios of input data from the spreadsheet and comparing the shadow model's computed results with the spreadsheet's output.

In the terms defined in this paper, the shadow model may be interpreted as an interface paradigm that differs from the cell-matrix paradigm and is tailored to serve as a better affordance for certain spreadsheet tasks. Ettema et al note several beneficial effects of working with the shadow model. A partial list of those benefits includes: clean separation of data and calculations; plain language variable names instead of alphanumeric cell-references; and clear access to the logic underlying a spreadsheet that is often otherwise difficult to discern. They also point out that incremental development of models is much easier in the shadow model than in a spreadsheet, because adjustments and supplements are often difficult or risky to incorporate in spreadsheets.

In fact, Ettema et al find these benefits so compelling that they ask 'Should spreadsheets be used at all' in a very prominent place in their paper. It would seem that the proper response to that question should be 'yes', if only because spreadsheets are extremely well suited for their original, pre-electronic, purpose: immediate presentation of data in tables. But this leads to another question: why should one be content for the shadow model to remain in the shadows? Why not keep the spreadsheet for data-presentation, but adopt the modelling language as an affordance paradigm for spreadsheet modelling, creation and comprehension?

Model Master, developed by Jocelyn Paine [Paine, 2001], takes a step in that direction. Model Master uses a text-based language for programming spreadsheets, which the Model Master compiler converts into actual spreadsheets. The syntax of Model Master, however, resembles that of object-oriented programming languages, which it intentionally mimics. It cannot be expected that the vast majority of business users of spreadsheets will feel at ease working with what to them will look like computer coding. Atebion [Atebion, 2005], another tool utilising an innovative interface, uses a mix of visual programming and natural English. [Nardi and Miller, 1990] note in a study they conducted of spreadsheet users that "the key to understanding non-programmers' interaction with computers is to recognise that non-programmers are not simply under-skilled programmers who need assistance learning the complexities of programming. Rather, they are not programmers at all. They are business professionals … whose jobs involve computational tasks".

## 3  DESIRED FEATURES OF A NEW PARADIGM

Following upon the above, we may now form a list of desired features for a new affordance paradigm for the various actions related to spreadsheets:

1. *A modelling language for the purposes of describing, comprehending and auditing the model expressed by a spreadsheet, whilst keeping the traditional spreadsheet itself for data-presentation.*

    1.1. The modelling language ought to be close to natural human language, whilst striving to avoid being so broad that it would permit confusing ambiguity.
    1.2. Ideally it should be indistinguishable from the language people use to describe business models when writing on white-boards or in text-books. It should be possible for a person with a reasonably general business education to

comprehend a written model directly, and to learn how to compose a written model rapidly. This would allow a design document to serve double duty as a programming tool and human readable documentation of the model or goal of the spreadsheet.
   1.3. It should be tolerant of variations in self-expression when those variations are non-ambiguous.
   1.4. The modelling language should be platform-independent – it should be possible to use the modelling language to generate spreadsheets in Excel, or StarOffice, or any other spreadsheet format.

2. *A generator that translates 'time-series models' – meaning models composed of variables and formulae spread over several time-periods, which constitute the bulk of business models – into spreadsheets.*

   2.1. The generator ought to free users as much as possible from repetitious and tedious tasks.
   2.2. The generated spreadsheet should not contain column-and-row cell-references. The generator should be clever enough automatically to implement named references.
   2.3. Being software, the generator, in going from model to spreadsheet, can provide much value-added over hand-crafted spreadsheets – for example, automatically separating data-only variables from calculated variables, presenting trees of dependency relationships between variables, analysing sensitivity rankings, and so forth.

3. *Separation of the conceptual model from the presentation structure of the spreadsheets.*

   3.1. It should be possible to define look-and-feel and general presentation structures separately from the model language.
   3.2. This can enable users to create spreadsheets that are uniform in look-and-feel and presentation structure simply by defining these parameters once, freeing them to concentrate on the important details of their models without the distractions of look-and-feel details.

The benefits that can be attained from software that can achieve these broad aims should be clear:

1) **Reduction of errors**. The best preventative of errors is a structural arrangement that reduces their likelihood of occurrence. The elimination of cut-and-paste actions from cell to cell, obscure cell-references and structural column-and-row alterations in the modelling of spreadsheets can be expected to translate into a reduction of entire classes of now common errors. Expressing models in plain language will also facilitate comprehension of formulae on the part of spreadsheet composers.
2) **Ease of auditing**. Reading a human-language description of the model underlying the spreadsheet affords immediate comprehension and can guide auditors to spot assumptions and formulae requiring special attention. The automatic insertion of named-references in cell formulae affords easier cell-by-cell inspection than alphanumeric references.

3) **Ease of documentation and maintenance**. Spreadsheets do not afford documentation. Text files do. A verbal spreadsheet model passed from one user to another, with documentation, is easier to maintain that a manually-created spreadsheet.
4) **Corporate-wide uniformity and re-use**. Utilising this interface paradigm can contribute to attaining the so-far mostly elusive goal of corporate-wide uniformity in spreadsheet lay-out styles and modelling methodologies. From the perspective of spreadsheet presentation, corporations can adopt uniform standards regarding matters such as left-to-right and top-to-bottom ordering of formulae, colour schemes, the location of variables requiring data-entry versus calculated data, and so forth, leaving only the modelling details to be changed as needed by individual projects and employees. With models separated from lay-out, it may be easier to institute corporate-wide instruction and uniformity in modelling methodology, and inculcate modelling cultures that focus on the most important elements of constructing financial models. Corporations can also develop and maintain libraries of formulae that can readily be re-used in models.
5) **Reduction of Sarbanes-Oxley Risk**. Given the provisions of the Sarbanes-Oxley Act, officers of corporations who cannot attest to the fact that they have reviewed financial statements, assured that they contain no untrue or misleading statements, and instituted internal controls to ensure the integrity of corporate financial statements may be liable for serious penalties. A modelling-language centred approach to generating spreadsheets can do much to mitigate this risk, in several ways. Simply having the ability to read a plain-language description of the model underlying a spreadsheet grants executives the ability to attest they have clear insight into the financial assumptions underlying corporate activities. Because the generated spreadsheets are separate from the verbal model, they can be locked from unauthorised editing, with cells calculated by formulae locked against over-writing.

## 4    AN IMPLEMENTATION: SUNSIGHT MODELLER$^{TM}$

Sunsight Modeller$^{TM}$ 1.0, produced by Inrise Financials, Inc., is a software system that implements the desired affordance paradigm for business-oriented spreadsheet creation, auditing and maintenance that has been put forward in this paper. It is intended for multiple uses, including scenario modelling, optimisation, research and calculation. We present here a very brief description of the main points of Sunsight Modeller$^{TM}$ 1.0 and some of its features.

### 4.1    Business Algebra Modelling Language

In order to use Sunsight Modeller$^{TM}$ 1.0 to create a spreadsheet, a user writes out what may be termed a 'Business Algebra Model', which is essentially a plain-language description of what the generated spreadsheet ought to contain. The Business Algebra Model may be composed in Microsoft Word or indeed virtually any text editor. The user would normally go through the following steps:

1. *Define the 'time-frame' which will appear in the spreadsheet.* This requires expressing, in words, the time period length, the number of periods and when the first period begins. A time frame might therefore look like:

Each period is one year.
The number of periods is 7.
The first period starts in 2005.

2. *Optionally define 'outline categories'*. Business models frequently require dividing variables amongst several categories. For example, sales numbers may be gathered separately based on products and markets, and then summed for presenting general totals. Sunsight Modeller<sup>TM</sup> 1.0 enables one to generate spreadsheets with these sort of category breakdowns very easily, and automatically sets outlines and calculates summed 'roll-ups' over the categories. The category hierarchies need be written only once, and after that only the title of each hierarchy has to be referenced in reports in order to have every variable broken down into the elements of the categories.

This done by writing out 'Categories:' and then writing an outline representing the hierarchies of categories, with the top or 'title element' of each hierarchy followed by an equals sign.

Here is an example

Categories:

Markets =
  1  North America
    1.1  Canada
    1.2  United States
  2  European Union
    2.1  United Kingdom
    2.2  France
Products =
  1  Standard
  2  Advanced

3. *Define reports and business drivers*. In the course of preparing a spreadsheet for presentation or financial analysis, one may generally want to present the data within separate 'reports'. For example, there may be a report which is a profit and loss statement, a report detailing return on assets, and a third report presenting cash flow analysis. Sunsight Modeller<sup>TM</sup> 1.0 enables one to divide business driver variables amongst such reports (note that a variable may appear in several reports). Each report may have its variables outlined according to the category hierarchies optionally defined above. Sunsight Modeller<sup>TM</sup> 1.0 will also automatically separate variables requiring data input from variables that are calculated from the data in other variables.

If categories have been defined and one wishes to have the report on the spreadsheet appear with any or all of the category elements in the outline, the next line should have the words 'Breakdown by' appear in it, followed by the titles of the category hierarchies to be included, which each hierarchy title separated by a comma. For example, a category breakdown might be written as 'Breakdown by Products, Markets'. After that, the formulae are written directly, with each formula separated from the next one by a new-line. Formulae in Sunsight Modeller<sup>TM</sup> 1.0 must have the assignment variable appearing to the left of the equal sign.

## 4.2 Sample Business Algebra Model

Each period is one year.
The number of periods is 3.
The first period starts on 2005.

Categories:
Markets =
  1  North America
    1.1  Canada
    1.2  United States
  2  European Union
    2.1  United Kingdom
    2.2  France
Products =
  1  Standard
  2  Advanced

Report: Profit And Loss
Breakdown by Markets
#====================================================
Gross Profit = Turnover – Cost of Sales

Operating Profit = Gross Profit – Selling and Administrative Expenses

Profit Before Taxes =  Operating Profit + Other Income – Interest

Profit = Profit Before Taxes – Taxes

Cost of Goods Sold = Labour + Raw Materials

Selling and Administrative Expenses = Selling and Distributions + Administrative Expenses

Report: Liquidity Analysis
Breakdown by Markets
#====================================================
Current Ratio  = Current Assets / Current Liabilities

Cash Ratio = (Cash + Short Term Investments)/Current Liabilities

Operating Cash Flow Ratio = Cash Flow from Operations / Current Liabilities

## 4.3 Sample Generated Spreadsheet

The sample Business Algebra Model shown above in section 1.5 generates, by way of Sunsight Modeller$^{TM}$, a full spreadsheet. We present here two screen captures of the automatically generated Excel spreadsheet derived from that sample:

The first sheet demonstrates the automatic separation of data assumptions variables from calculated variables (the numerical data in this sheet were entered manually, but all the cells not coloured in yellow were computer generated):

| | Assumptions for Markets | Years | | |
|---|---|---|---|---|
| | | 2005 | 2006 | 2007 |
| | | 0 | 1 | 2 |
| 7 | North America | | | |
| 53 | European Union | | | |
| 54 | United Kingdom | | | |
| 55 | Cost of Sales | 27,095 | 28,450 | 29,872 |
| 56 | Turnover | 51,514 | 54,090 | 56,794 |
| 57 | Selling and Distributions | 12,605 | 13,235 | 13,897 |
| 58 | Administrative Expenses | 6,868 | 7,211 | 7,572 |
| 59 | Other Income | 12 | 13 | 13 |
| 60 | Interest | 1,646 | 1,728 | 1,815 |
| 61 | Taxes | 343 | 360 | 378 |
| 62 | Raw Materials | 19,924 | 20,920 | 21,966 |
| 63 | Labour | 7,171 | 7,530 | 7,906 |
| 64 | Current Assets | 13,401 | 14,071 | 14,775 |
| 65 | Current Liabilities | 19,955 | 20,953 | 22,000 |
| 66 | Short Term Investments | 1,226 | 1,287 | 1,352 |
| 67 | Cash | 1,678 | 1,762 | 1,850 |
| 68 | Cash Flow from Operations | 7,497 | 7,872 | 8,265 |
| 69 | France | | | |
| 84 | European Union.All Markets | | | |
| 99 | All Markets | | | |

The second sheet presents the calculations of profit and loss. Everything appearing in this sheet was automatically generated by the software, with no manual human input – indeed, the cells are security-locked to prevent tampering. The reader is invited to note the readability of the formula generated for the highlighted cell, as compared to a formula depending upon alphanumeric cell-references:

## 5 CONCLUSION

We have shown that consideration of spreadsheet user-interaction paradigms that differ from the standard 'cell-matrix' paradigm that has dominated the electronic spreadsheet over the past two and a half decades can provide significant benefits in multiple categories for business-orientated spreadsheet users. We have also suggested a list of desired features for a new 'affordance paradigm' for spreadsheet creation, and presented a working software system built upon the principles of such an alternative paradigm.

Further studies researching and quantifying how different spreadsheet user-interaction paradigms can impact usability, ease-of-use, error-rates, precision, readability, comprehension, maintenance, security and ease of auditing of spreadsheets are clearly called for. It also remains to be seen to what extent a new spreadsheet paradigm can overcome barriers to widespread adoption amongst users who have gotten used to the existing paradigm for decades. This latter challenge will likely require a combination of utility value and capitalising on network effects by integrating as readily as possible to already widespread tools and programs.